The Fermi Paradox revisited: Technosignatures and the Contact Era

Amri Wandel, Racah Institute of Physics, The Hebrew University of Jerusalem

Abstract

A new solution to the Fermi Paradox is presented: probes or visits from putative alien civilizations have very low probability until a civilization reaches a certain age (called the Contact Era) after the onset of radio communications. If biotic planets are common, putative advanced civilizations may preferentially send probes to planets with technosignatures, such as radio broadcastings. The contact probability is defined as the chance to find a nearby civilization located close enough, so that it could have detected the earliest radio emissions (the "radiosphere") and sent a probe that would reach the Solar System at present. It is found that the current contact probability for Earth is very low, unless civilizations are extremely abundant. Since the radiosphere expands with time, so does the contact probability. The "Contact Era" is defined as the time (since the onset of radio transmissions) at which the contact probability becomes of order unity. At that time alien probes (or messages) become more likely. Unless civilizations are highly abundant, the Contact Era is shown to be of the order of a few hundred to a few thousand years and may be applied not only to physical probes but also to transmissions (i.e. SETI). Consequently, it is shown that civilizations are unlikely to be able to inter-communicate, unless their communicative lifetime is at least a few thousand years.

1. Introduction

One of the possible resolutions of the well-known Fermi Paradox, "Where are they?" suggests that Earth may be the only civilization, since there is no evidence for extraterrestrial visits. This hypothesis builds on the "colonization scenario": advanced civilizations capable of space-travel at a velocity of a few percent of the speed of light would eventually colonize the whole Galaxy within a few million years (Hart 1975; Cox 1976). Such a scenario requires a long-lasting social commitment and billions of expeditions or probes. More recent solutions of the Fermi Paradox suggest that life is rare, e.g. the Rare Earth hypothesis (Ward and Brownlee 2000), that intelligent life is rare (e.g. Sandberg, Drexler and Ord 2018) or that advanced technological civilizations with spaceflight are rare (e.g. the Great Filter hypothesis of Hanson, 1998). Many more resolutions to the Fermi Paradox have been discussed by Zuckerman and Hart (1995) and more recently by Webb (2015). Recently some works and reports consider the hypothesis that alien civilizations have already visited the Solar System. In particular, the peculiar properties of the extrasolar object Oumuamua inspired the space probe hypothesis (Bialy and Loeb 2018), along with less exotic "natural" explanations (e.g. Bannister et al. 2019). Also the report on Unidentified Aerial Phenomena (UAP; USA National Intelligence report 2021) have triggered initiatives like project Galileo (Harvard U. news 2021) and arose questions about their origin and nature. However, recent scientific articles (e.g. Curran 2021; Zuckerman 2021) as well as popular ones (e.g. Kolbert

2021) refute the hypothesis of Oumouamoua being an alien artificial object. In this work we suggest that alien interpretations of UAPs and extrasolar objects are improbable at present, but may become more probable in a few hundred or thousand years.

Although we have no information about the capabilities of putative alien civilizations, it is reasonable to assume that they are not unlimited and that civilizations with limited resources would choose planets with special properties as targets to their probes. Presuming biological life is common, Earth's biosignature would not be outstanding. In other words, to extraterrestrial civilizations would not consider Earth as special, since there are probably many biotic planets closer to them. This is supported by the high abundance of habitable-zone planets (Dressing and Charbonneau 2013; 2015). However, it can be shown that unless civilizations are highly abundant, the probability to find a neighbor civilization that Earth is one of its few nearest habitable planets is very low (Wandel 2021; 2022).

If biotic planets are common (the Copernican hypothesis, W15, Wandel 2022), a special property distinguishing Earth could be its technosignature, e.g. artificial radio transmissions. Indeed, intelligence and technosignature may be much rarer than biotic signs, i.a. because of the temporal argument: while planets can remain biotic for billions of years and show biosignatures for a long epoch, civilizations showing technosignatures may exist for a much shorter time (however, see Wright et al. 2022 for a scenario, in which technological civilization outlive their biological ancestors). In particular, Earth's technosignature is still undetectable to alien civilizations unless sufficiently nearby: the information about Earth being home to a technological civilization has not yet reached beyond a hundred light-years. This information is carried by Earth's "radiosphere", the expanding envelope of "intelligent" radio signals from Earth that began to spread in space less than a century ago, with the first shortwave radio transmissions and the use of radar before World War II.

While this work is mainly concerned with physical probes, a similar criterion was used by Wandel (2015, hereafter W15) for alien transmissions intentionally aimed at Earth, arguing that such transmissions must originate from a civilization nearer than ~50 light years, since the envelope of these signals (Earth's radiosphere) has expanded to less than 100 light years from Earth. The present work applies this argument to physical objects, alien spaceships or probes, which makes a stronger case than radio signals, as the energetic cost of interstellar flight is much higher than electromagnetic transmissions and its speed is lower.

The basic assumption of this work is the hypothesis that alien civilizations do not bother to explore merely biotic planets, so that the radiosphere criterion becomes the main trigger for alien civilizations to send interstellar probes (in sections 3-4 we extend this argument to directed transmissions). This leads to the new concepts of contact probability and Contact Era, which allow a quantitative approach to the Fermi Paradox.

2. The radiosphere technosignature criterion

If biotic planets are highly abundant, they may be less attractive targets for physical exploration and colonization. One possible solution to the Fermi Paradox is that putative civilizations do not have enough resources or do not persist long enough to colonize or explore all the planetary systems in the Galaxy, or even all habitable planets. If their resources are not unlimited, civilizations would not send probes to every biotic planet in their galactic neighborhood, as there may be millions or billions of such planets (depending on the distance defined by "neighborhood"). This argument may be supported by our own experience – the only civilization we know so far: even if we had the resources, humanity would not consider sending Rosetta-like missions to every single asteroid in the Asteroid Belt or to every comet in the Oort Cloud. We assume that civilizations would restrict their efforts to a subset of planets that look particularly attractive, e.g. having a technosignature, in the civilizations' galactic neighborhood. Applying this assumption to Earth, we assess the probability that an alien civilization is near enough to have detected Earth's radio signals and consequently have sent probes or messages that could reach us at present. Obviously this probability depends on the abundance of civilizations. Estimates of the number of civilizations in the Galaxy vary greatly (e.g. Livio 1999; Snyder-Beattie 2021). We find that the probability that such a civilization would be close enough to Earth is extremely small, unless civilizations are highly abundant.

Assuming that a precondition for the evolution of technology is life, the number of planets with technology within a distance $a$ from Earth is given by the product of the biotic and intelligence fractions by the density of stars:

$$N_i(a) = 40 \, (f_b/0.1) \, (f_i/0.01) \, (n^*/0.01 \text{ ly}^{-3}) \, (a/100 \text{ ly})^3, \qquad (1)$$

where $n^*$ is the local space density of stars suitable for the evolution of life on planets around them. Following the definition of the parameters in the Drake equation, $f_b$ is the fraction of biotic planets and $f_i$ is the fraction of biotic planets which evolve intelligent species. Assuming that all stellar types from G-M can support the evolution of life, the space density of suitable stars, $n^*$, may be close to the total local stellar density, $0.004$ ly$^{-3}$ (Gregersen 2010). It may be significantly less, if certain kinds of stars are excluded. Taking into account the geometry and dimensions of galactic disk, eq. 1 is valid for $a<h_G$, where $h_G \approx 500$ ly is the scale height of the stellar galactic population; for $a>h_G$ the last term in eq. 1 takes the form $\sim a^2 h_G$.

The Kepler mission has shown that ~10-50% of the stars host a small rocky planet within their Habitable Zone (e.g. Dressing and Charbonneau 2013; 2015). Assuming a large fraction of those planets are biotic, we normalize $f_b$ to $0.1$. Of course, $f_b$ may be much smaller, if biotic planets turn out to be rare. Similarly, we (arbitrarily) normalize $f_i$ to $0.01$. As noted, also $n^*$ could effectively be smaller than the nominal stellar density, if certain types of stars are excluded. In particular, if M-dwarfs, which consist 75% of all stars, are excluded, as their planets may be considered less suitable to host life (e.g. Bochanski 2010; Gale and Wandel 2017; Lingam and Loeb 2019), the effective $n^*$ decreases by a factor of ~5. Excluding binary stars as well, as the orbits of their planets may be unstable, would lower $n^*$ by another factor of ~2. Fig. 1 shows the number of expected planets with civilizations (presumable that could send probes) as a function of the distance from Earth, for a couple of values of the product $f_b f_i$ and two choices for $n^*$: including binaries and M-dwarfs (0.01; solid curves) and excluding those groups (0.001, dashed curves).

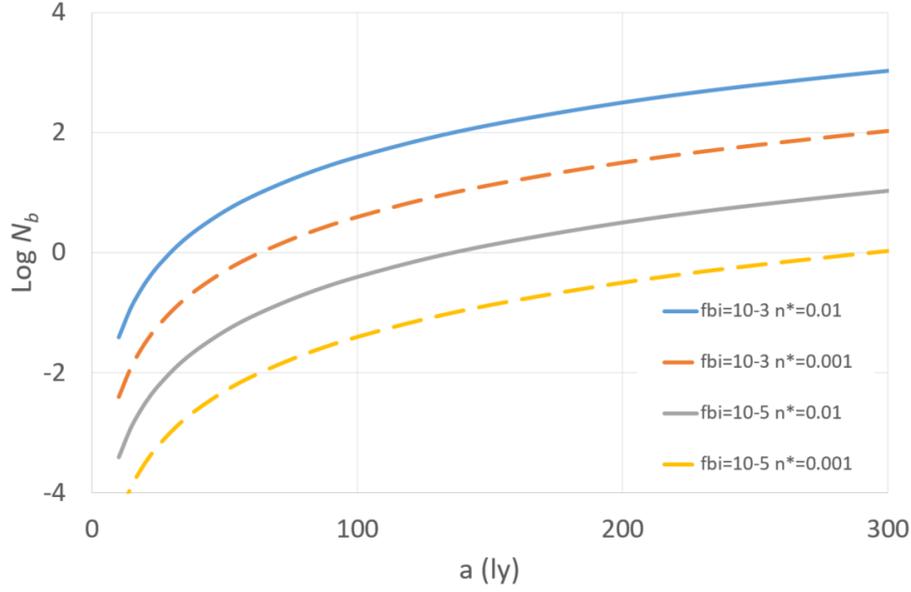

Fig. 1. The number of expected civilizations as a function of the distance from Earth. Solid curves include all stellar types ($n^*=0.01$), dashed ones exclude M-dwarfs and binaries ($n^*=0.001$). $f_{bi}$ in the legend represents the product of $f_b$ and $f_i$, the biotic and intelligence Drake parameters.

The probability to find a civilization within a given distance from another one (e.g. ours) depends on the average distance between neighbor civilizations. Assuming $N$ civilizations are evenly scattered in a thin galactic disc of diameter $D \sim 10^5$ light-years ly, a simple geometric calculation gives for the average distance between neighbor civilizations $d(N) \approx D\, N^{-1/2}$ assuming $d$ is larger than the thickness of the disk, which is the case if $N \sim < 10^5$. Else if $d$ is comparable or smaller than the scale height of the stellar population ($h_G \sim 500$ ly), which is the case if $N >\sim 10^5$, the average distance between neighbor civilizations is $d \sim (D^2 h_G/N)^{1/3}$. Of course, civilizations like stars, may be unevenly distributed. Unlike stars, we do not know how civilizations are distributed, so uniform distribution is a good assumption. Following these assumptions, the probability $p$ of finding a civilization within a distance $a<d$ is $p \sim (a/d)^2$ for $N<10^5$, while for $N<10^5$ we get $p \sim (a/d)^3$, which gives

$$p(N,a)=N(a/D)^2 \qquad\qquad N<10^5 \qquad\qquad (2)$$

$$p \sim N a^3/(D^2 h_G). \qquad\qquad N>10^5$$

The first short-wave radio transmissions that could penetrate the ionosphere and leak to space were broadcasted in the 30's, less than 100 years ago. Therefore, the maximal distance of a civilization that could detect Earth's radiosphere and eventually send back a message that would reach Earth at present is approximately 50 ly. For a probe of speed $v=\beta c$, the maximal distance for which this scenario could work is

$$a=100/(1+1/\beta) \text{ ly}. \qquad\qquad (3)$$

E.g. at a velocity of *0.2c*, a probe could reach the Solar System at present (e.g. in 2017, the year Oumeamea has been detected) only from civilizations nearer than 16 ly. Combining eqs. 2 and 3 gives

$$\log p = \log N - 6.0 - 2 \log (1+1/\beta) \quad\quad N<10^5 \quad\quad (4)$$

$$\log p = \log N - 7.0 - 3 \log (1+1/\beta). \quad\quad N>10^5$$

Note that setting $\beta=1$ in eq. 3 gives a constraint on the distance of a civilization from which alien transmissions could have reached Earth, $a<50$ ly. Substituting in eq. 4 gives a constraint on the probability to receive an alien transmission by present, $p<10^{-7}N$. This could explain the absence of detections in the SETI project, the so-called "Big Silence". This result is further elaborated in section 4.

Fig. 2 shows the probability $p(N,\beta)$ of finding a civilization within the maximal "radiosphere probe distance" given by eq. 3, with the lower horizontal axis indicating the probe's velocity and the upper one the distance. E.g., for $N<\sim 10,000$ and $\beta<\sim 0.1$, the probability is $<\sim 10^{-5}$. Note that as the distance of the nearest exoplanet, Proxima Centauri b is ~4 ly, eq. 3 gives a lower limit on the speed of alien probes that could reach Earth at present, $\beta>0.04$.

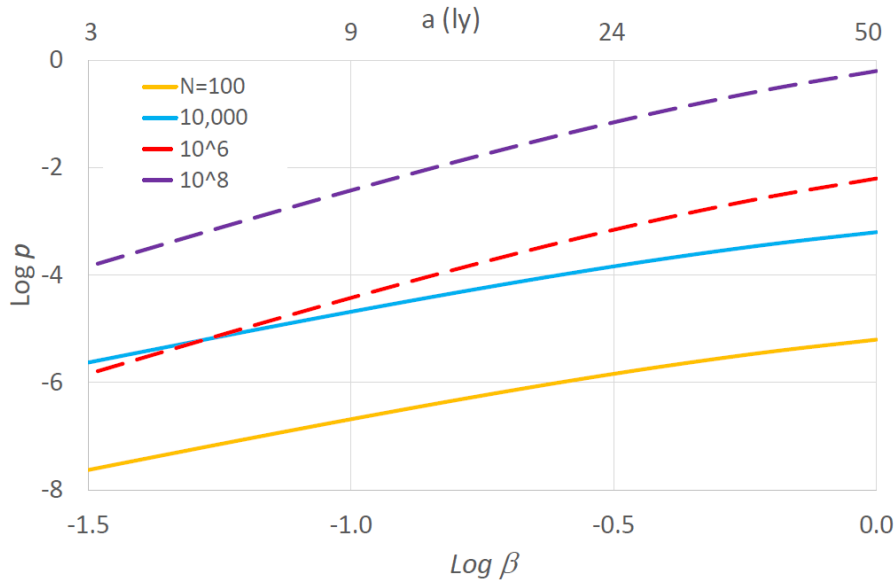

Fig. 2. The probability (log *p*) to find a civilization at a distance *a* (upper horizontal axis), from which a probe of speed $\beta=v/c$ (lower horizontal axis) could reach Earth 100 years after the first transmissions from Earth. The solid curves are for *N=100* (lower curve) and 10,000 (upper curve) civilizations assumed to be uniformly scattered in the Galaxy. The dashed curves are for $N=10^6$ (lower) and $10^8$ (upper) civilizations.

Fig. 3 shows iso-probability contours in the $N-\beta$-plane. That is, the probability (as a function of $\beta$ and *N*, eqs. 4) to find a civilization within a distance *a*, from which a probe of speed $v=\beta c$ could reach Earth at present, having been sent when the first radio signals from Earth could have been detected.

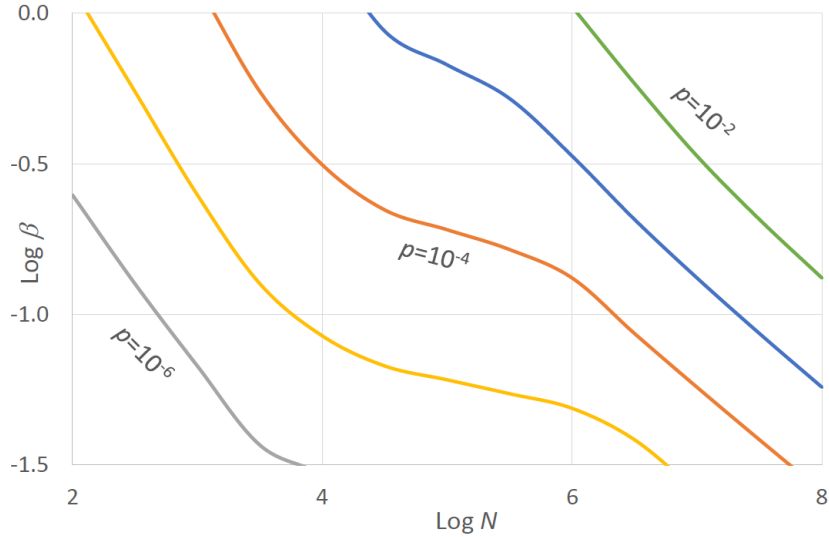

Fig. 3 Iso-probability contours of finding a civilization within the radiosphere probe-distance defined in the text, in the $N$–$\beta$ parameter plane.

The energy needed to accelerate even a relatively small probe to speeds close to the speed of light is enormous: for example, the energy required for accelerating a probe of 1000 kg to 0.1c is ~$10^{25}$ erg, more than $10^4$ times the global yearly energy production of Earth. Hence presumably, even if an advanced civilization is technologically capable of reaching velocities close to the speed of light, it may prefer to use lower speeds. As we have seen, for the (present) radiosphere probe criterion $v>0.04c$, but in the future Earth (or for older civilizations, see section 3) probe speeds might be lower.

In the case of radio-transmissions, advanced civilizations may install automatic radio beacons (Benford 2010), increasing the effective value of $N$ beyond the number of actual civilizations. However, this argument is less applicable to space probes, as launching probes from a beacon in response to signals is less feasible than sending radio signals.

3. The Contact Era

Extending the above analyses in time, we derive an age at which civilizations are more likely to be reached by probes or transmissions from neighbor civilizations. Since the radiosphere expands with time, we may write an expression for the maximal distance defined in eq. 3 (half the radius of the radiosphere),

$$a(t) \sim a_0\, (t/t_0),$$

where for the present Earth $a_0 = 50$ ly and $t_0 = 100$ years. Consequently, the probability for a probe (or an intentional directed transmission from an alien civilization) increases with time, as can be written combining the radiosphere size with eqs. 4:

$$p(t) \sim 6\cdot 10^{-6}\, N(t/t_0)^2 (1+1/\beta)^{-2} \qquad N<10^5 \qquad (5)$$

$$p(t) \sim 10^{-7} N(t/t_0)^3 (1+1/\beta)^{-3} \qquad N>10^5$$

Setting $p\sim1$ in eq. 5 and solving for $t$ gives a universal timescale $t_c$, which we call the Contact Age or the Contact Era, the age of a civilization (measured from the time it begins using radio communication) when receiving alien transmissions ($\beta=1$) or probes ($\beta<1$) becomes more likely:

$$t_c \sim 100{,}000\, N^{-1/2}(1+1/\beta) \text{ yr} \qquad N<10^5 \qquad (6)$$

$$t_c \sim 20{,}000\, N^{-1/3}(1+1/\beta) \text{ yr} \qquad N>10^5$$

Obviously, for two civilizations to inter-communicate (i.e. to exchange two-way messages) the Contact Age must be larger than their longevity as communicating civilizations. This condition can be shown to give a constraint on interstellar inter-communication, resulting in a lower limit on the longevity, below which two-way exchange of messages has a low probability (Wandel 2021;2022). In order to derive this result, we use the reduced form of the Drake equation (W15)

$$N \sim R^* f_b f_i L,$$

where $R^*\sim1$ is the product of the astronomical terms, $f_b$ and $f_i$ are the biological and intelligence factors defined above, and $L$ is the average longevity of communicative civilizations. The probability for a civilization to establish two-way communication with a neighbor civilization before it reaches the Contact Era is small. In other words, to be able to receive an answer to a message sent to another civilization, the longevity must be longer than the Contact Era: inter-civilization communication requires $t_c<L$.

This condition could apply also to receiving an intended alien signal (see section 4). As electromagnetic transmissions move at the speed of light, $\beta=1$ and the Contact Era for signals becomes

$$t_c \sim 200{,}000\, N^{-1/2} \text{ yr} \qquad N<10^5 \qquad (7)$$

$$t_c \sim 40{,}000\, N^{-1/3} \text{ yr}. \qquad N>10^5$$

For example, in the range $10^4<N<10^6$ the Contact Era starts at a time between 400-2000 years after the onset of radio communications.

Combining this expression with the reduced Drake equation gives a general constraint on the longevity (average duration of communicative civilizations),

$$L > \sim 3500\, (R^* f_b f_i)^{-1/3} \text{ yr} \qquad N<10^5 \qquad (8)$$

$$L > \sim 2800\, (R^* f_b f_i)^{-1/4} \text{ yr}. \qquad N>10^5$$

As $R^*\sim1$, the term in parenthesis is of order unity or smaller, hence eq. 8 gives a condition for two-way inter-civilization communication in the form of an absolute constraint (independent of other parameters) on the longevity, $L>\sim3000$ years. In the case that the abundance of civilizations is small ($f_i \ll 1$), the constraint becomes even stronger, e.g. for $f_i \sim 1\%$ $L>10{,}000$ yr.

Figs. 4 and 5 show the Contact Era as a function of *N and v/c,* respectively. This is the age of a civilization, measured from the time it begins to transmit, at which receiving probes (or beamed transmissions) from neighbor civilizations becomes more probable. Different curves refer to different values of *v/c* (Fig. 4) and *N* (Fig. 5). If there are less than a million civilizations in the Galaxy, the Contact Era for transmissions *($\beta$=1)* is longer than ~400 years. If we exclude non-intended beamed signals (such as radar) and beacons, this may suggest a new explanation to the lack of SETI-detections (the so-called Great Silence). For probes the contact timescale is yet longer by a factor of ~$1/\beta$, and beacons and unintended signals do not apply. Hence, as noted above, the Contact Era formalism may be considered as new solution to the Fermi Paradox*.*

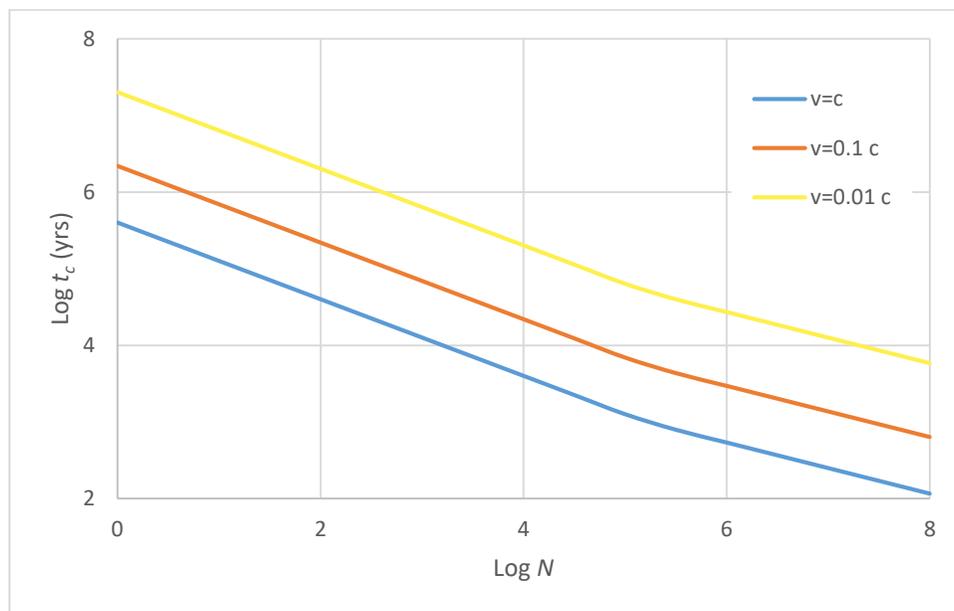

Fig. 4. The Contact timescale defining the epoch in the life of a civilization when receiving probes or directed transmissions becomes more probable, as a function of the number of civilizations in the Galaxy. The three curves refer to different probe velocities: *v=c* (the lower curve; this curve also applies to radio transmissions), *v=0.1c* (middle) and *v=0.01c* (upper).

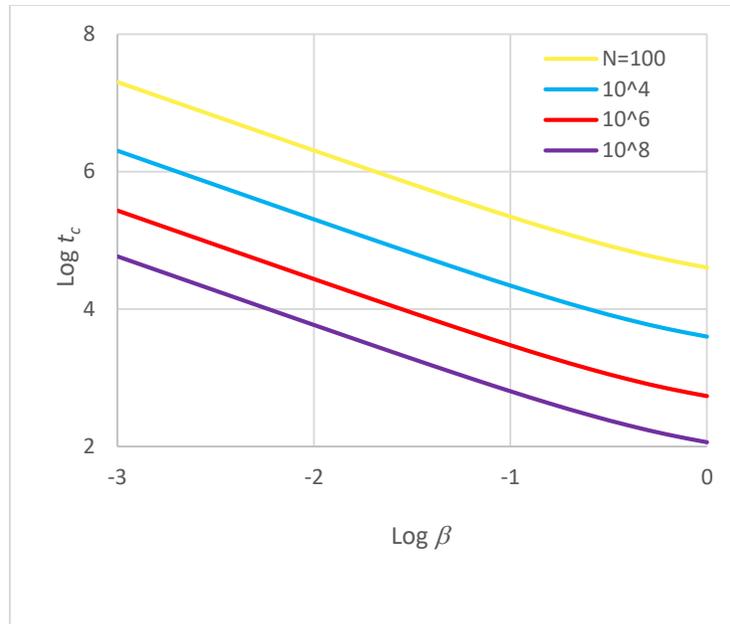

Fig. 5. The Contact timescale as a function of probe velocity. Different curves refer to the number of civilizations in the Galaxy: 100 (upper curve), $10^4$, $10^6$ and $10^8$ (lower curve).

4. SETI

As leakage signals may be too faint (W15) or time-limited, e.g. civilizations would eventually switch their internal communication from omnidirectional radio transmissions to a radio-quiet technology like fiber optics, the Search of extra-Terrestrial Intelligence (SETI) may have to focus on directed alien transmissions. If, according to the Copernican hypothesis (W15 and the Introduction) biotic planets are common, such transmissions could hardly be attracted by the biosignature of Earth but are more likely to be triggered by Earth's technosignature, namely by the radiosphere. Isotropic signals leaking from local radio communication do not need to be much stronger than on Earth. Such leakage signals could be detected by our present technology from a distance of a few tens of light years (Sullivan at al. 1978; Loeb add Zaldarriaga 2007). Of course, alien detectors could be more sensitive than ours, which is taken into account by introducing the isotropic signal distance defined below.

Assuming that alien isotropic transmissions are significantly less detectable than directed ones, nearby or beamed alien signals directed at Earth would be more easily detectable by our present technology. Under this assumption the Contact Era (eqs. 7) may explain the lack of detections by SETI. An alternative to beamed intended transmissions from alien civilizations could be interstellar lighthouse beacons or AI-based civilizations (Gale, Wandel and Hills 2019). Such beacons could transmit isotropically and independently of detecting Earth's technosignature, hence they would not be subject to the radiosphere criterion. However also beacons are likely to have a limited range, so they too are subject to the isotropic signal analyses derived below.

In order to assess the range of validity of these arguments, we define the isotropic signal probability $p_{is}$. Let us assume that $N$ civilizations are evenly scattered in the Galaxy with an average isotropic transmission power which our current sensitivity can detect from a distance $d_{is}$. Taking into account the limited duration of the isotropic transmission phase, assuming the average longevity of a civilization with such isotropic transmissions is $L_{is}$, the distance to the furthest isotropically transmitting civilization that can be detected is

$$d_{is,L} = \text{Min}\,(d_{is},\, cL_{is})$$

In order to estimate the probability of finding such a civilization near enough to Earth, in analogy to eqs. 4, we define the "isotropic probability" as

$$\log p_{is} = \log N - 6.6 + \log (d_{is,L}/100 \text{ ly}) \qquad N<10^5 \qquad (9)$$

$$\log p_{is} = \log N - 7.9 + \log (d_{is,L}/100 \text{ ly}). \qquad N>10^5$$

For example, if the average isotropic detection distance is 100 ly and there are 10,000 civilizations with $L_{is}$>100 years, the isotropic probability is 0.002. A million civilizations with an average isotropic detection distance of 1000 ly and a longevity >1000 years would give an isotropic probability of ~0.1. For the beacon scenario, if each one of 10,000 civilizations posts beacons within a radius of 1000 ly from its home planet, with an operating time >1000 years, the isotropic probability would be 0.025.

for $p_{is}$ <1. Eqs. 9 show that this condition is satisfied when the product

$$Nd_{is,L} \sim <10^{10}.$$

As this condition covers most of the likely parameter space, we conclude that the Contact Era formalism may be applied also to transmissions and SETI, unless civilizations are extremely abundant (e.g. $N>10^8$) and their isotropic transmission power is huge (e.g. 10,000 times more than that of Earth).

5. Discussion

If biotic planets are so abundant that habitability and life alone do not provide a sufficient motivation for alien interstellar exploration, planets with technosignatures may attract alien civilizations to send probes. The probability that a civilization is located close enough to Earth, to detect our radiosphere and send a space probe that would reach the Solar System at present is found to be very small, unless civilizations are extremely abundant ($10^8$ or more civilizations in the Galaxy). Civilizations may be attracted to send probes at the Solar System by Earth's radiosphere, but unless they are highly abundant, our radiosphere has not yet reached the nearest civilization. As it expands, the probability that Earth's radiosphere will engulf an alien civilization increases with time. Alien probes would become more likely at the Contact Era, which depends on the assumed probe speed and civilization abundance. For fiducial values it varies between hundreds and thousands of years. These arguments, if valid, may resolve the Fermi Paradox, and may predict a timescale when alien probes or directed transmissions become more likely.

Alien civilizations nearer than ~50 ly could have detected Earth's radiosphere and may have sent directed radio signals which could be detected by SETI (W15). This argument may be used as a weak Fermi-SETI-paradox: since we have not yet detected such signals, perhaps there are no civilizations nearer than ~50 ly. This translates into a constraint on the number of civilizations in the Galaxy, $N<10^7$. Of course, there are many caveats (e.g. Brin 1983): the civilization may not be communicative, or directed signals may have been sent when we were not listening etc. Similarly, if a civilization sufficiently nearby has sent probes to the Solar System, its eventual communication with the probes would be directed at Earth and may be detected by SETI. Also in this case similar caveats apply, e.g. the probes may be autonomous or communications may be seldom. It is shown that for directed signals the Contact Era may apply not only to physical probes and alien visits but also to alien intended transmissions, and eventually explain the lack of detections by SETI.

Acknowledgements: the author thanks an anonymous reviwer for many valuable comments that helped to improve the article.